\documentclass[12pt]{article}
\usepackage{amssymb,epsfig} 
\def\ads{{\rm AdS}_{1,d-1}} \def\cm{{\rm CM}_{1,d-2}}
\def\so{SO($2,d\!-\!1$)}
  \def\RR{\mathbb{R}}

\normalbaselines 
\begin{document} 
\begin{center}
\vspace*{1.0cm}

{\LARGE{\bf A Proof of the AdS-CFT Correspondence%
\large\footnote{Talk given
    at ``Quantum Theory and Symmetries'', Goslar (D), July 1999}}}

\vskip 1.5cm

{\large {\bf K.-H. Rehren }} 

\vskip 0.5 cm 

Institut f\"ur Theoretische Physik \\ Universit\"at G\"ottingen \\ 
D-37073 G\"ottingen (Germany)  
\end{center}

\vspace{1 cm}

\begin{abstract}
\noindent The AdS-CFT correspondence is established as a re-assignment
of localization to the observables which is consistent with locality and
covariance.  
\end{abstract}

\vspace{1 cm} 

\section{Introduction}
The ``holographic principle'', or ``AdS-CFT correspondence'' has been 
considered as an amazing prediction on the basis of string theory,
D-branes, and supergravity, setting off an enormous research activity. 
It states that the ``degrees of freedom'' of a quantum field theory in
anti-deSitter (AdS) space-time of dimension $d$ can be completely
identified with the degrees of freedom of a conformal quantum field theory 
in Minkowski space-time of one dimension less. It was first formulated
for supergravity in 5 (+ 5 compactified) dimensions and super
Yang-Mills theory in 4 dimensions by J. Maldacena \cite{Mal}, and was then
conjectured by E. Witten to hold as a rather general principle \cite{Wit}. 

The holographic principle arises by a confluence of several ideas arising
from string theory and from gravity. It was discussed long ago by
't Hooft \cite{Hoo} that the degrees of freedom of quantum fields {\em above}
a black hole horizon are counted by the Bekenstein entropy {\em of} the
horizon. In string theory, the horizons of certain extremal black hole
solutions to classical supergravity with AdS geometry are
considered as D3-branes. These appear as the
supporting space-time for a 1+3-dimensional superconformal Yang-Mills
theory, whose state space should contain that of the supergravity theory. 

The AdS-CFT correspondence, yet, is not a feature of string theory. 
Although string theory happened to play a prominent role in its
discovery, the conjecture itself is a statement about ordinary 
QFT, and its validity should be discussed in the framework of
ordinary QFT.   

It is well known that the AdS-CFT correspondence admits no simple
identification of the respective quantum fields on the two space-time
manifolds involved. It has also by now become a familiar conception
with many of us that the quantum fields proper are somewhat ambiguous
entities in QFT while the invariant entities are the
local algebras they generate \cite{Haa}. If we agree to take this 
idea seriously, a precise formulation of the conjectured correspondence
is possible without any recourse to string theory or to perturbative
supergravity, and a rigorous proof is astonishingly simple \cite{Reh}. 

The formulation of a QFT should be such that it admits the unambiguous
and relativistically invariant computation of all the physical
quantities (such as masses and cross sections) which specify the theory.  
It has been demonstrated that this possibility is guaranteed if, in a 
representation of positive energy, only the localization of each observable is 
known (see \cite{Buc}). The covariant assignment of localizations to the
observables (or, what amounts to the same, the specification of all
observables which are localized in a given region) provides the
physical interpretation and therefore determines the theory. 
This formulation of a QFT does not rely on any idea of quantization of
some classical physics.

The idea for the AdS-CFT correspondence is that the same observables
can be given different assignments of localizations (in different
space-times) in a covariant way, and are thus interpreted differently 
as a QFT in $d$ dimensions or a conformal QFT in $d-1$ dimensions. We
shall establish the conjecture by showing that any covariant
assignment of a localization in one space-time to the observables
gives rise to another such assignment in the other space-time, and
vice versa. But we shall see that it is not always adequate to think
of localization in terms of smearing of fields.

So what is the role of string theory? 
It might turn out on the long run, that string theory is just an 
ingenious device to produce nonperturbative QFT's, and might not at 
all exceed the realm of ordinary QFT. This must not be a 
reason to be disappointed but may be celebrated as a scheme which
circumvents or soothens the notorious difficulties attached to
Lagrangean perturbation theory. If this is correct, then it is no
surprise if ``stringy'' pictures lead to statements about QFT. 

While locality has a central position in QFT, the intuition prevails 
that string theory is ``less local'' than ordinary QFT, and hence is a new
type of theory. But quite to the contrary, string theory is even more
local than is necessary in QFT: namely, if one accepts that (free)
fields carrying different inner degrees of freedom commute
irrespective of their localization, then, treating transversal string
excitations as inner degrees of freedom, they may cause excited
string fields 
to commute even at time-like separation. This crude idea has been
elaborated with some rigor for free strings \cite{Dim}. I personally
take it as another hint that, as the dust settles, string theory might
bring us back to QFT, hopefully at an advanced level of insight.

\section{Algebraic holography}

We formulate the AdS-CFT conjecture as the assertion that the
local QFT's on AdS space-time $\ads$ are in 1:1 correspondence with
the conformally invariant local QFT's on compactified Minkowski
space-time $\cm$. Corresponding theories have the same state space.

AdS space-time is usually described as the hypersurface 
$\{x:x_0^2-\vec x^2+x_d^2=R^2\}$ in an ambient Minkowski space
with two time directions (or rather, its quotient by identification of
antipodal points $x,-x$). The embedded space has only one intrinsic
time direction and a global time orientation. A serious reason
for unease is that it has closed time-like curves. Yet, the
Klein-Gordon field constructed in \cite{Fro} is, for quantized values
of the mass, a perfectly sensible QFT on AdS space-time.  

QFT on AdS space-time has also been studied in \cite{BFS}. No
a priori conflict with the general framework of QFT was found, except that
causal influences (signalled by non-commutation of local observables)
can propagate only along geodesics. This feature (well known also from
chiral observables in 2D conformal QFT's) must presumably be
interpreted as the absence of proper interaction. Thus, interacting QFT's
will, like the Klein-Gordon field of generic mass \cite{Fro}, live on
a covering space. 

One may also regard AdS space-time $\ads$ as a ``cosmological
deformation'' of Minkowski space-time with the maximal number of isometric
symmetries. The symmetry group (AdS group) is the Lorentz group \so\ of the
ambient space, the boosts in the second time direction acting as
deformed translations.  

The very same group, \so, is also the symmetry group of conformal QFT
in $d-1$ dimensions, that is on compactified Minkowski space-time
$\cm$. In fact, the manifold $\cm$ coincides with the conformal
boundary of AdS space-time (defined by the conformal structure of
$\ads$ induced by the AdS metric), and its conformal structure
coincides with the one inherited by restriction. The AdS group
preserves the boundary and acts on it like the conformal group.

This observation is pivotal for the asserted 1:1 identification of
QFT's: the Hilbert space and the representation of
the group \so\ are the same for both theories. What differs is the
interpretation of the group: e.g., the Hamiltonian of the
AdS group (the rotation between the two ambient time
directions) corresponds to a periodic subgroup of the conformal group
generated by time translations and time-like special conformal
transformations. Space-like AdS translations correspond to
dilatation subgroups of the conformal group. 

The set of observables, considered as operators on the common Hilbert
space, is also the same for corresponding theories. What differs is
the assignment of a localization to a given observable. The same
operator is said to be localized in a suitable region $X\subset\ads$ 
as an observable of the AdS theory, and it is said to be
localized in another region $Y\subset\cm$ as an observable of the
conformal theory.    

For this reinterpretation to work, it must be compatible with
covariance and locality, and this is the nontrivial issue. The group \so\ 
must act geometrically in both interpretations, and hence the sets of 
regions $X$ and $Y$ above must be related by a bijective
correspondence which respects the action of the group. Furthermore,
this bijection must map causal complements with respect to one
geometry into causal complements with respect to the other geometry
since the geometric notion of causal independence is coded
algebraically by local commutativity.

These constraints already fix the bijection between regions $X$ in
AdS space-time and regions $Y$ in conformal Minkowski space-time of
one dimension less. Here it is: 

The basic regions in $\ads$ are wedge-like regions which
arise as the connected components of the intersection of AdS
space-time with the ambient space region $\{x: x^1 > \vert x^0 \vert\}$,
and all \so\ transforms thereof. Every such intersection provides a
pair of wedge regions which turn out to be causally complementary in the
sense that they cannot be connected by a causal geodesic.

The basic regions in $\cm$ are nonempty intersections of a forward and
a backward light-cone, called double-cones. Each wedge region in
$\ads$ intersects the boundary in a double-cone in $\cm$. 
This yields a bijection between wedge regions in $\ads$ 
and double-cone regions in $\cm$ which respects the action of the
group \so\ by construction, and causal complements by inspection. 

Sofar, the discussion is entirely geometric. The algebraic part is
very simple: We declare an operator which, as an observable of the
boundary theory, is localized in some double-cone, to be
localized, as an observable of the AdS theory, in the
associated wedge region, or vice versa. A minute's thought brings
about that this prescription yields a 1:1 correspondence between
local QFT's on AdS space-time $\ads$ and local QFT's on conformal 
Minkowski space-time $\cm$ \cite{Reh}. The associated theories
have the same Hilbert space and the same representation of the group
\so, while their respective physical interpretations are different.

One issue remains to be discussed, the positivity of the energy
spectrum. As we mentioned before, the two interpretations go along
with different Hamiltonians. By a standard argument, it is well known
that if one of these Hamiltonians has positive spectrum, then so does
the other. Yet, the spectra are very different! The AdS Hamiltonian
has discrete spectrum due to periodicity in time, while the boundary
Hamiltonian has conformally invariant, hence continuous spectrum.

We shall discuss that the notion of sharp localization acquires a very
different meaning in corresponding theories. While sharp
boundary localization obviously corresponds to localization at
space-like infinity of AdS space-time, sharp AdS localization 
turns out to have a very delicate meaning in terms of
boundary localization.

\section{Examples}
The proof sketched above is a structural proof for a structural
statement. It does not tell us which particular AdS theory is
associated with which particular conformal QFT. This has to be
discussed case by case, as will be exemplified below. 

Our proof also doesn't tell us which observables are localized in
bounded regions of AdS space-time. The standard procedure is to say
that an observable is localized in an arbitrary AdS region $X$, if it
is localized in all wedge regions which contain $X$. The according 
algebraic determination of localization in AdS double-cone regions
yields the following general structural results \cite{Reh}.  

In $d\geq 1+2$, if there are double-cone localized AdS observables, the
boundary theory must violate the additivity property that the
observables localized in small double-cones covering the space-like
basis of a large double-cone generate the observables localized in
the large double-cone. While this additivity should always hold for
theories generated by gauge-invariant Wightman fields, its violation
seems characteristic for non-abelian gauge theories, as Wilson
loops cannot be expressed in terms of point-like gauge invariant
quantities. Thus, AdS theories which are described in terms of proper
local fields \cite{Fro} and which consequently possess observables localized in
bounded regions, should correspond to gauge-type conformal boundary theories. 

Conversely, a boundary theory which is additive in the above sense
must correspond to a QFT on AdS space-time which does not possess
local fields. To prevent confuion: the latter still has as many
wedge-localized observables 
as there are observables in the boundary theory, which however cannot
be ``detached'' from infinity. Impossibility of point-like localization
does not mean that the theory is non-local, since observables
localized in causally complementary wedges do commute as they
should. Topological field theories (such as Chern-Simons with Wilson  
lines attached to the boundary as wedge-localized observables) come to
one's mind \cite{Wit}.  

The situation is much more favorable in $d=1+1$. Then $\cm=S^1$ and the
boundary theory is a chiral conformal QFT. The Penrose diagram of
two-dimensional AdS space-time is the strip $\RR\times (0,\pi)$ with
points $(t,x)\sim (t+\pi,\pi-x)$ identified. This is in fact a
M\"obius strip. Light rays are $45^\circ$ lines. A wedge region is a
space-like triangular region enclosed by a future and a past directed 
light ray emanating from a point in the interior of the strip, and the
associated boundary ``double-cone'' is the interval cut out of the
boundary $S^1$ by this wedge.    

A double-cone in the M\"obius strip is the intersection of two
wedges which cut out of the boundary $S^1$ two intervals 
$I_1,I_2$ which overlap at both ends: $I_1 \cap I_2 = J_1 \cup J_2$.

An AdS observable in $d=1+1$ is thus localized in a double-cone if as a chiral 
observable it is at the same time localized in {\em both} the large
intervals $I_1,I_2$. This is of course true for the chiral observables 
localized in $J_1$ or $J_2$, but in general there will be more than
these. The additional AdS double-cone observables remain broadly
localized (if considered as boundary observables) even if the
double-cone, and hence the intervals $J_1,J_2$ are small. We shall
see examples of such observables below. This again does not mean any
non-locality of AdS double-cone observables, since they do commute
whenever two double-cones sit within causally complementary wedges. It
only means that their description in terms of boundary fields may be
non-local.   

We present an example in $d=1+1$: a massless conserved vector current
$j^\mu$ on AdS space-time represented by the M\"obius strip. 
Its equations of motion are solved in the plane by
$j^0(t,x)=j_R(t-x)+j_L(t+x)$ and $j^1(t,x)=j_R(t-x)-j_L(t+x)$. 
Restricting this solution to the strip $\RR\times (0,\pi)$ and
requiring it to respect the identification of points 
$(t,x)\sim (t+\pi,\pi-x)$ amounts to put $j_L=j_R\equiv j$ with period
$2\pi$. Canonical quantization is achieved by representing
$[j(u),j(u')]=i\delta'(u-u')$ on a Fock space.

This is indeed a U(1) current, the simplest conformal QFT on
$S^1$. But its degrees of freedom are redistributed (through the
combinations $j^\mu$) over the M\"obius strip.  

We now compute observables localized in an AdS double-cone (giving rise to
intervals $I_1 \cap I_2=J_1\cup J_2$ as before) in the interior of the
strip. Typical boundary observables localized in an interval $I$ are
of the form $W(f)=\exp ij(f)$ where $f$ is a periodic smearing
function which is constant outside the interval $I$. Adding a constant
$c$ to $f$ is immaterial since the charge operator $\int j(u) du$ is 
a multiple $q$ of unity in every irreducible representation,
so $W(f+c)=e^{icq}W(f)$. Now, consider a smearing function which has
constant but different values on both gaps between $J_1,J_2$. 
Then $W(f)$ is localized as a boundary observable in both
intervals $I_1,I_2$, but it is neither localized in $J_1$ nor in $J_2$, nor
is it generated by such observables. As an AdS observable,
$W(f)$ is localized in a double-cone, and operators of this form
generate all observables in the double-cone. Suitably regularized
limits of $W(f)$ even yield point-like local fields on the M\"obius
strip $\phi(t,x)=\exp i\alpha\int_{t-x}^{t+x}j(u)du$.
 
As the size of a double-cone well within the strip shrinks, the
correlations of the boundary observables within the intervals
$J_1,J_2$ disappear, giving rise to two decoupled (chiral) current 
algebras, one for $J_1$ and one for $J_2$, among the AdS
observables of the double-cone. The additional observables of the form
$W(f)$ as discussed above behave in this limit as vertex operators   
$E_\alpha(t+x) \otimes E_{-\alpha}(t-x)$ carrying opposite left and right
chiral charge. Thus, locally the QFT on AdS space-time looks
like the bosonic sector of the Thirring (``U(1) WZW'') model.

\end{document}